\documentclass[aps,prl,reprint,groupedaddress]{revtex4-1}
\usepackage{graphicx, color, soul}
\usepackage{siunitx}
\usepackage{amsmath}
\usepackage{color,soul,url}

\begin{document}
\title{Broadband topological slow light through higher momentum-space winding}

\author{Jonathan Guglielmon$^{1}$ and Mikael C. Rechtsman$^{1}$}
\affiliation{$^{1}$Department of Physics, The Pennsylvania State University, University Park, PA 16802, USA}

\date{\today}

\begin{abstract}
Slow-light waveguides can strongly enhance light-matter interaction, but suffer from narrow bandwidth, increased backscattering, and Anderson localization. Edge states in photonic topological insulators resist backscattering and localization, but typically cross the bandgap over a single Brillouin zone, meaning that slow group velocity implies narrow-band operation. Here we show theoretically that this can be circumvented via an edge termination that causes the edge state to wind many times around the Brillouin zone, making it both slow and broadband.
\end{abstract}

\maketitle

When propagating through a medium, a pulse of light can travel with a group velocity that is much slower than its vacuum value \cite{Hau1999,Ku2004,Vlasov2005}. This phenomenon of slow light has been extensively studied due to its potential for applications ranging from optical buffers to enhanced light-matter interactions (and thus nonlinearity) \cite{Hau1999,Xu2000,Soljacic2002,Povinelli2005,McMillan2006,Krauss2008,Baba2008,khurgin2008slow}. A well-known problem encountered in slow-light systems arises from fabrication imperfections: as one decreases the group velocity of the light, it becomes increasingly sensitive to disorder, leading to significant backscattering, loss, and localization \cite{Melati2014}. In recent years, significant research effort has been dedicated to studying and realizing photonic topological insulators \cite{Haldane2008,wang2009observation,Hafezi2011,Umucalilar2011,Fang2012,rechtsman2013photonic,hafezi2013imaging,khanikaev_2013,Peano2016,Fang2017,ozawa2018}.  These systems possess chiral edge states that resist backscattering and localization in the presence of disorder. They, therefore, constitute natural candidates for generating robust slow light.  Additionally, in contrast to typical slow-light systems, which require special designs to avoid reflections from the slow-light interface \cite{Schulz2010}, chiral edge states will automatically exhibit complete transmission between topological fast-light and topological slow-light regions, independent of the slow-light group index and the details of the interface.

\begin{figure*}
\centering
\includegraphics[width=\textwidth]{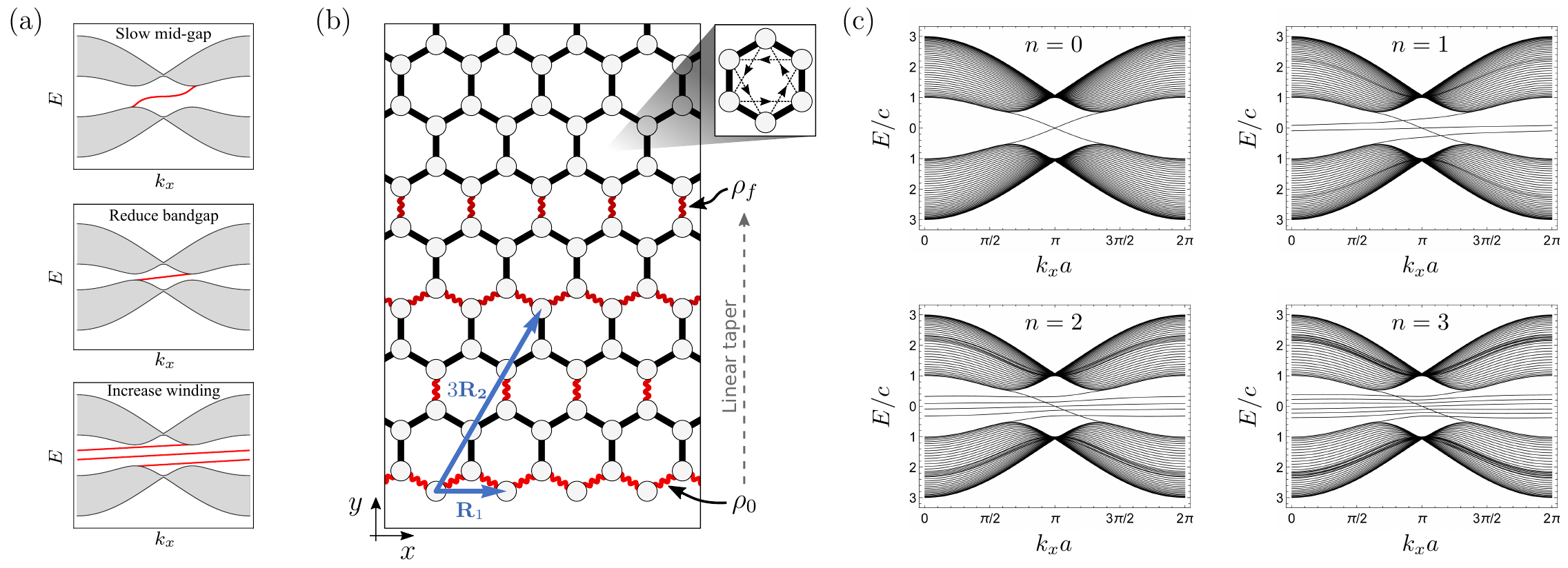}
\caption{\label{fig_pattern} Panel (a) compares band structures for different methods of generating a slow topological edge state. The first two methods slow the edge state over a narrow range of energies while the third method yields a slow edge state with a large bandwidth. Panels (b) and (c) show how the third method can be implemented in the Haldane model. Panel (b) illustrates the edge termination, with wavy red lines indicating reduced nearest-neighbor couplings. These couplings are reduced by a factor that begins at $\rho_0$ at the edge and linearly tapers to a final value $\rho_f$. Up to these factors, the nearest-neighbor coupling pattern repeats $n$ times under translation by $3\mathbf{R}_2$ before terminating into the bulk (shown is the case $n=2$). Second-neighbor couplings (not shown) are rescaled using a simple linear taper (see text). As indicated in the upper right corner of the panel, the direction of positive phase hopping for the second-neighbor couplings is taken to be counter-clockwise.  Panel (c) shows the resulting band structures. As $n$ is increased, the edge state winds an increasing number of times around the Brillouin zone as it crosses the bandgap. For clarity, we have only structured the lower edge so that only the bottom-localized edge mode exhibits an increased winding.}
\end{figure*}

In many applications, the usefulness of a slow-light system depends crucially on its bandwidth which, ideally, should be large so that the light can be slowed over a large range of frequencies \cite{Baba2008}. In topological systems, a typical edge termination\textemdash such as a zig-zag edge of a honeycomb lattice\textemdash produces an edge mode that crosses the bulk bandgap over a single Brillouin zone. As a result, reducing the group velocity of the edge mode requires either slowing the mode only in the vicinity of a given energy (e.g., mid-gap) \cite{Yang2013} or reducing the bandgap (see Fig. \ref{fig_pattern} (a)). In both cases, the reduced group velocity comes at the expense of bandwidth. Additionally, in the latter case, the reduced bandgap means that the existence of the edge mode will be more sensitive to disorder since disorder strong enough to close the bandgap can induce a topological phase transition. 

In this letter, we demonstrate that, by engineering the edge termination, a topological edge mode can be made to wind many times around the Brillouin zone as it crosses the bandgap, thereby generating a slow edge mode over a large range of frequencies. The number of times the edge mode winds is determined by the depth of the modification of the edge termination measured into the bulk. In the direction parallel to the edge, the termination does not expand the size of unit cell and, therefore, generates multiple windings in a manner distinct from simple band folding. Since the mode is slowed without reducing the bulk bandgap, its existence remains protected against strong disorder. The ability to slow the mode without reducing its bandwidth is enabled by the fundamentally 2D nature of the system, as different frequencies reside at different depths in the structure. As a result, the minimal group velocity attainable at a fixed bandwidth is determined by the system size in the direction orthogonal to the direction of propagation. In contrast to typical topological slow-light systems, where the 2D footprint of the bulk may be viewed as a drawback, our proposed structure makes use of this region to enable wideband operation. 

Designing the edge termination requires specifying a set of parameters that can be tuned along the edge. The implementation of the photonic topological insulator\textemdash for example, whether it is realized using magneto-optics \cite{Haldane2008,wang2009observation}, modulated resonators \cite{Fang2012,Minkov2016}, or optomechanics \cite{Fang2017}\textemdash will determine this set of tunable degrees of freedom. The central observation of this letter, however, is not restricted to a specific implementation but rather is generally applicable to systems containing chiral edge states. We, therefore, avoid assuming a specific photonic implementation and instead conceptually demonstrate the features of our proposal in the Haldane model \cite{haldane1988model}.

We begin with the Haldane model defined on a honeycomb lattice with real first-neighbor couplings and complex second-neighbor couplings. We set the inversion symmetry breaking mass to $M=0$ and the phases for the second-neighbor couplings to $\phi = \pi/2$, with positive phases assigned to counter-clockwise hopping (see Fig. \ref{fig_pattern}). We denote the magnitudes of the first-neighbor and second-neighbor couplings by $c$ and $c'$, respectively. For definiteness, we take $c' =\frac{1}{10}c$. We denote the lattice constant by $a$ and the lattice vectors by $\mathbf{R}_1 = a(1,0)$ and $\mathbf{R}_2 = a(\frac{1}{2},\frac{\sqrt{3}}{2})$. The non-zero value for $\phi$ places the system in a topologically non-trivial phase with Chern number $C=1$ so that opening the boundaries of the system produces a topological edge mode that crosses the bulk bandgap as the momentum parallel to the edge is swept across the Brillouin zone. We will consider a horizontal strip geometry that is periodic in the $x$-direction and finite in the $y$-direction, terminated on zig-zag edges. We choose coordinates such that the lowest site resides at $y=0$. We then modify the Hamiltonian in the vicinity of the edge by tailoring the couplings to control the behavior of the edge mode. For simplicity, we will leave the upper edge of the structure unchanged and only introduce modifications to the lower edge.

On the lower edge, we pattern the nearest-neighbor couplings in the way illustrated by Fig. \ref{fig_pattern}(b). In particular, we reduce a subset of the couplings\textemdash those indicated in the figure by wavy red lines\textemdash by a factor $\rho(y)$ that depends on the height, $y$, of the link center for the neighbor pair (i.e., $c\to \rho(y) c$). The pattern alternately rescales horizontal and vertical couplings, while interspersing regions in which the couplings are left unchanged. The pattern maintains full $\mathbf{R}_1$ periodicity. Up to the values of the rescaling factors, the pattern repeats after translation by $\mathbf{S} = 3\hspace{1pt}\mathbf{R}_2$. After $n$ repetitions along $\mathbf{S}$, the pattern terminates into the bulk of the standard Haldane model (i.e., with all nearest-neighbor couplings set to $c$). The resulting edge termination, therefore, extends to a depth of $n\mathbf{S}$ into the bulk. As we will see shortly, the value of $n$ determines the number of times the edge mode winds around the Brillouin zone. 

We choose the rescaling function, $\rho(y)$, so that couplings residing close to the edge are reduced more than couplings residing deeper in the bulk. We take the rescaling factor to begin at a value $\rho_0$ on the edge and increase linearly to a final value $\rho_f$ before terminating into the bulk. Defining $y_0$ and $y_f$ as the $y$-coordinates of the link centers for the first and last rescaled nearest-neighbor couplings, respectively, we take ${\rho(y) = \left(\frac{\rho_f - \rho_0}{y_f-y_0} \right)(y-y_0) + \rho_0}$. Nearest-neighbor couplings residing beyond $y_f$ are not rescaled and are set to their bulk value $c$. 

Similarly, we rescale next-neighbor couplings by a factor $\rho'(y)$ which begins at the edge at $\rho_0'$, ends in the bulk at $\rho_f'$, and is linearly tapered in between. In contrast to the nearest-neighbor couplings, every next-neighbor coupling near the edge is rescaled (i.e., no couplings are skipped, as they are for the nearest-neighbor pattern). The first rescaled next-neighbor coupling resides at $y=0$ and the final rescaled next-neighbor coupling resides at $y=n S_y$ (i.e., the depth reached after $n$ translations along $\mathbf{S}=(S_x,S_y)$). Hence, the rescaling function for the next-neighbor couplings is given by ${\rho'(y) = \left(\frac{\rho_f' - \rho_0'}{nS_y} \right)y + \rho'_0}$. Next-neighbor couplings with $y>nS_y$ are set to their bulk value $c'$. Tuning the parameters $(\rho_0,\rho_f)$ and $(\rho_0',\rho_f')$ allows us to control the dispersion of the edge mode. In the remainder of this letter, we will set $(\rho_0,\rho_f) = (0.05,0.28)$ and $(\rho_0',\rho_f') = (0.15,1.00)$. These values were chosen to produce a simple edge dispersion exhibiting clear additional windings around the Brillouin zone. 

\begin{figure}
\centering
\includegraphics[width=0.95\columnwidth]{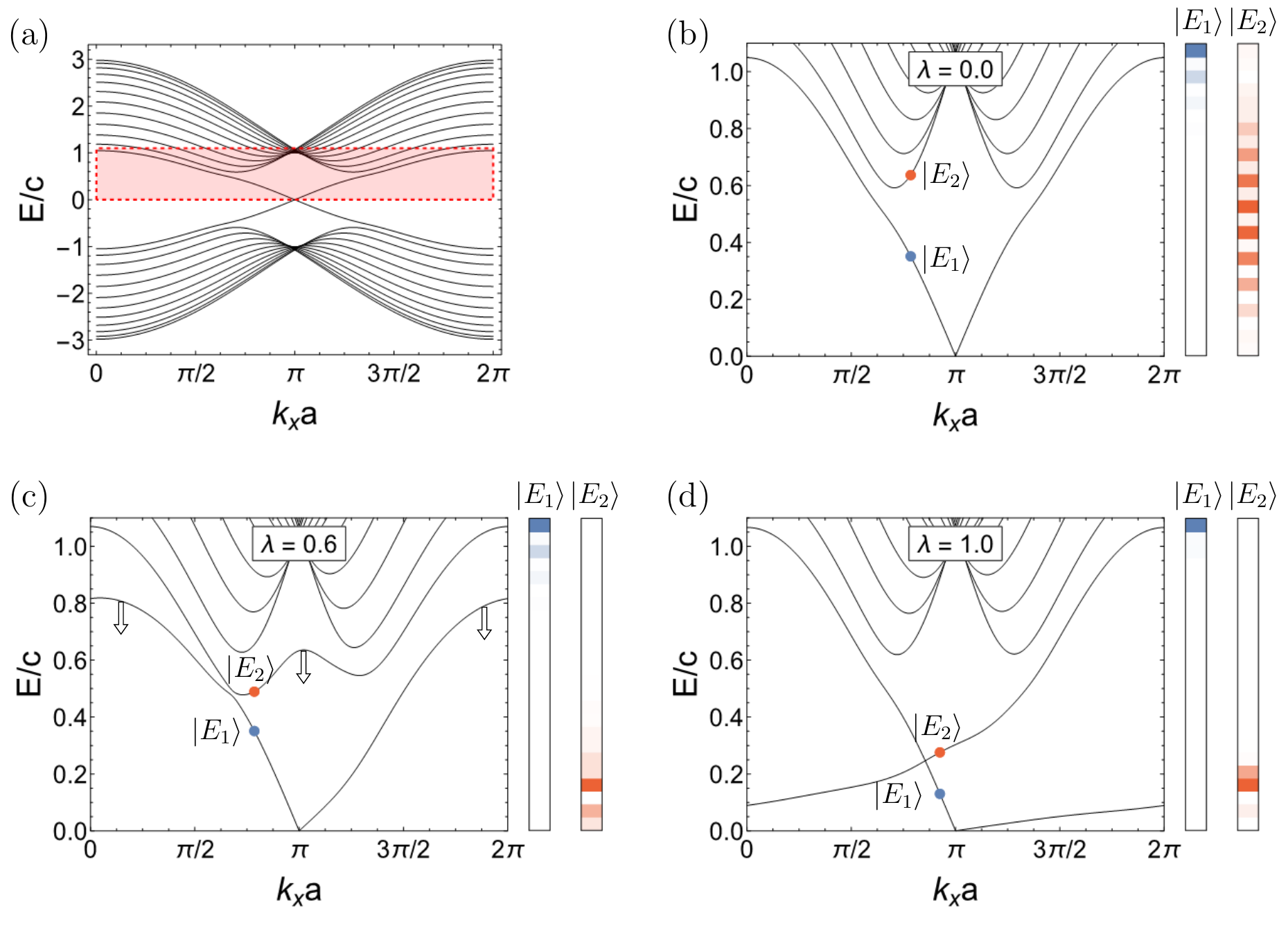}
\caption{\label{fig_hybridize}Conversion of bulk modes to edge modes under a smooth transformation between the $n=0$ and $n=1$ edge terminations. Panel (a) shows the $n=0$ band structure, highlighting in the red box the region that is magnified in the neighboring plots. Panels (b)--(d) show the transformation of the bands as $\lambda$ is swept from $0$ to $1$. A small system size is used to clearly distinguish distinct bands. The two panels located to the right of each band structure show the $y$-dependence of the eigenstate intensity profiles (over a single strip-geometry unit cell) for the eigenstates $|E_1\rangle$ and $|E_2\rangle$ highlighted in the band structure. As $\lambda$ is increased, a bulk mode is pulled into the gap and becomes localized on the bottom edge, allowing it to cross the top-localized mode without hybridizing. A similar process occurs simultaneously for a bulk band residing in the region $E<0$ (not shown). The resulting bottom-localized edge mode winds around the Brillouin zone two additional times as it crosses the gap.}
\end{figure}

With the rescaling functions set as described above, the number of repetitions, $n$, determines the number of times the edge mode winds around the Brillouin zone as it crosses the bandgap. A standard zig-zag edge is reproduced by taking $n=0$. Sending $n$ to $n+1$ causes the edge mode to wind two additional times (i.e., $\Delta k_x \to \Delta k_x  + \frac{4\pi}{a}$) around the Brillouin zone as it crosses the bandgap. The resulting band structures for $n=0,1,2,3$ are shown in Fig. \ref{fig_pattern}(c). 

For the $n=0$ edge termination (a standard zig-zag edge), the edge modes associated with the upper and lower edges together form a continuously connected pair of bands. To understand how this pair of bands can develop additional windings, we study how the edge mode transforms under a smooth interpolation between the $n=0$ and $n=1$ terminations. We define $H_0(k_x)$ and $H_1(k_x)$ to be the Bloch Hamiltonians for the $n=0$ and $n=1$ cases, respectively, and define a one-parameter family of Hamiltonians ${H_\lambda(k_x) = (1-\lambda)H_0(k_x) + \lambda H_1(k_x)}$ that smoothly interpolates between $H_0(k_x)$ and $H_1(k_x)$ as $\lambda$ is varied over the interval $[0,1]$. In Fig. \ref{fig_hybridize}, we show a magnified view of the resulting transformation. As $\lambda$ is increased, a bulk mode is pulled into the gap and approaches the edge mode localized on the top edge (Fig. \ref{fig_hybridize}(c)). Typically, these modes would exhibit an avoided crossing due to their spatial overlap. However, as the bulk mode is pulled into the bandgap, it becomes localized on the bottom edge, so that the overlap becomes exponentially suppressed in the system size, allowing it to cross the top-localized edge mode (Fig. \ref{fig_hybridize}(d)). A similar process simultaneously occurs for a bulk band residing below the bandgap. As a result, the edge mode acquires two additional windings around the Brillouin zone.

\begin{figure}[t!]
\centering
\includegraphics[width=0.95\columnwidth]{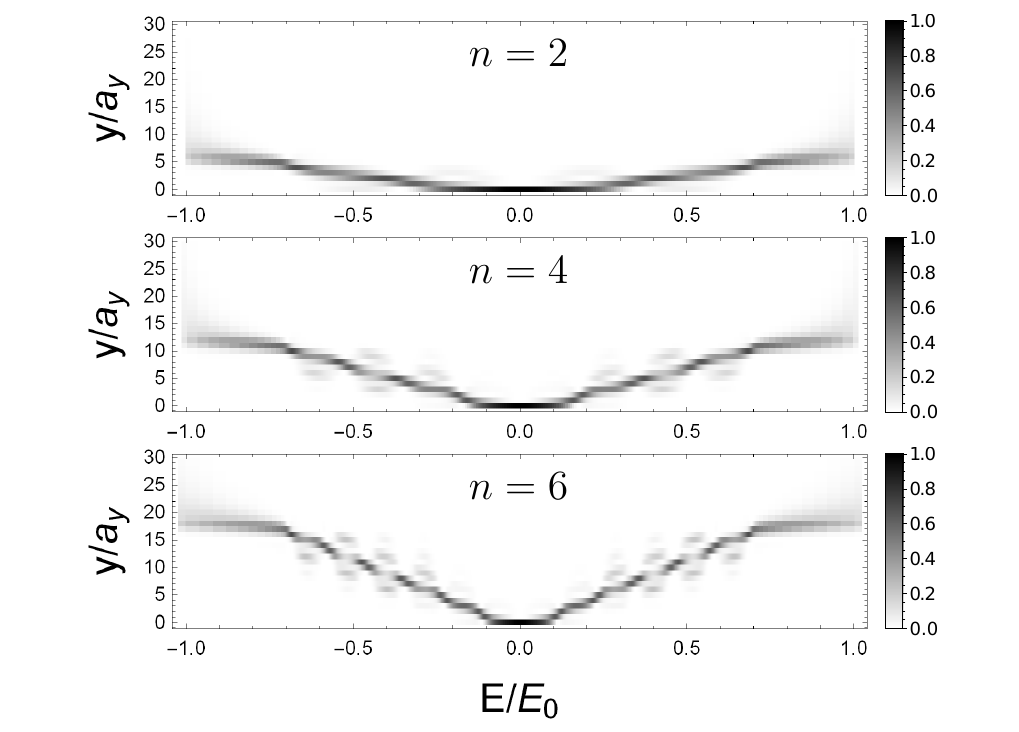}
\caption{\label{fig_edgestates} Edge mode eigenstate profile as a function of energy for the optimized edge terminations. Each constant $E$ slice shows the $y$-dependence of the intensity distribution (over a strip-geometry unit cell) of the edge state at energy $E$. Only the 30 sites nearest to the edge are shown. The $y$-coordinates are normalized to $a_y = \mathbf{R}_2^y$ and the energy $E$ is normalized to units of the bandgap so that $E/E_0 = \pm1$ correspond to the bulk band edges. Separate panels are shown for the optimized edge terminations defined by $n=2,4,6$. As the number of windings is increased, the edge mode utilizes degrees of freedom residing deeper in the bulk.}
\end{figure}

As the number of windings increases, the edge mode utilizes degrees of freedom that reside at increasing depths in the bulk. To study how the edge mode eigenstate profile varies as its energy is swept across the bandgap, we first optimize the coupling pattern to minimize variations in the group velocity (see Supplemental Material \cite{suppmat}) so that the bottom-localized edge mode crosses each energy in the bandgap exactly once and each energy is associated with a unique eigenstate (i.e., the optimization removes any non-monotonicity present in the edge bands and gives them nearly linear dispersion). Figure \ref{fig_edgestates} shows how the edge mode eigenstate profile changes as the energy, $E$, is swept across the gap. The intensity profile is shown over a single strip-geometry unit cell. For simplicity, the intensities associated with the two sublattices have been (additively) coarse-grained into a single intensity profile. At mid-gap, $E=0$, the edge mode resides at the very edge of the structure. Away from mid-gap, it moves deeper into the bulk, with increasing depths occupied by the mode as the number of windings is increased. Note, however, that even as the mode moves into the bulk, it maintains a small cross-sectional mode profile. This feature is important for achieving enhanced light-matter interactions in a slow-light system.

In typical slow-light systems, a reduction in the group velocity is accompanied by an increased sensitivity to fabrication imperfections so that light is more easily scattered and localized by disorder. Topological edge states, however, are known to resist localization and backscattering. To demonstrate how this feature applies to a topological slow-light device, we perform time-domain simulations in which we launch a narrow-band pulse into a disordered slow-light region. We perform these simulations both for a topological structure and for a topologically trivial 1D array. For the topological structure, we use the optimized edge terminations so that, in the absence of disorder, the edge band has nearly linear dispersion. For the trivial structure, we control the group velocity by varying the nearest-neighbor coupling.

\begin{figure}
\centering
\includegraphics[width=\columnwidth]{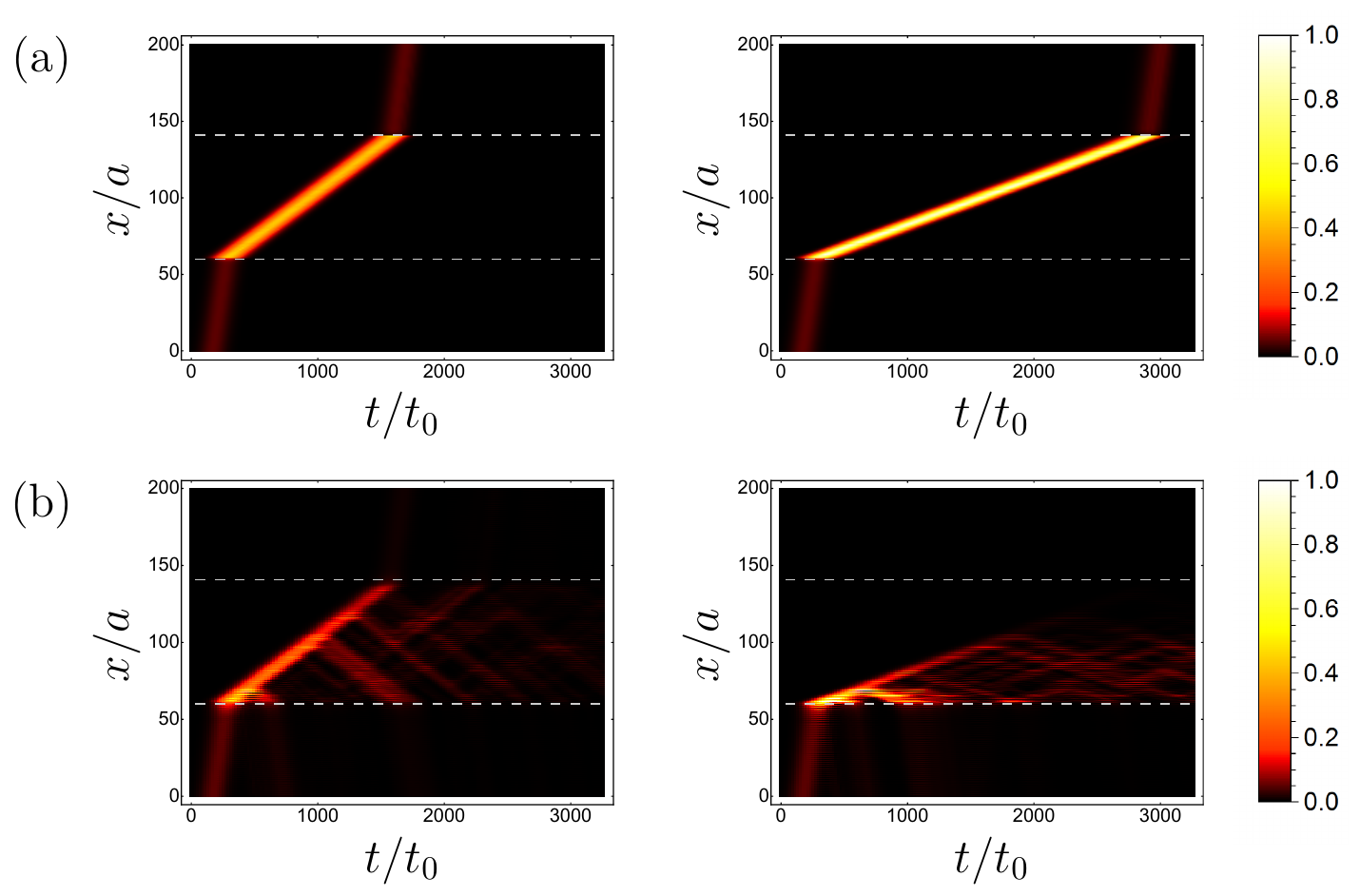}
\caption{\label{fig_disorder}Propagation of slow light in the presence of onsite disorder. A pulse initially traveling in a fast-light region enters and propagates through the slow-light region located between the dashed white lines. Row (a) shows the topological case using the optimized edge terminations. The left and right panels correspond to slowing factors of $v/v'=9$ and $v/v'= 18$, respectively. Each constant $t$ slice shows the intensity profile of the edge mode along the $x$-direction (the direction of propagation). The time coordinate, $t$, is scaled to units of the coupling time $t_0 = 1/c$. To show the pulse purely as a function of $x$, the intensities have been summed over the strip-geometry unit cell, collapsing the $y$-dependence of the pulse. Row (b) shows the corresponding results for a topologically trivial 1D system.}
\end{figure}

Defining $v$ and $v'$ as the fast-light and slow-light group velocities, respectively, we independently perform the simulations for $v/v' = 9$ and $v/v'= 18$ (corresponding to the optimized $n=1$ and $n=2$ edge terminations) using onsite disorder with a strength of $5\%$ of the bandgap of the topological structure. We describe these simulations in the Supplemental Material \cite{suppmat}. The results are shown in Fig. \ref{fig_disorder}. The topological structure exhibits a clear improvement, resisting the significant localization and backscattering that increase in severity for the trivial system as the group velocity is reduced. More generally, for a wideband pulse, the topological system will display a similar improvement, but the pulse shape will undergo distortion due to disorder-induced dispersion.

In conclusion, we have shown that, by increasing an edge state's momentum-space winding, it can be slowed without sacrificing bandwidth. For the resulting structures, the magnitude of the group velocity is decoupled from the size of the bandgap and from the periodicity along the propagation direction. The structures, therefore, circumvent the recently suggested limitation \cite{Minkov2018} that the strength of the topological protection derived from the size of the bandgap is tied to the maximal group index. 

For our proposed structure, the wideband nature of the edge state may be viewed as a natural consequence of slowing the mode in the presence of non-trivial topology: due to the non-zero Chern number, the edge state is required to fully traverse the bandgap so that, in response to a reduction of the group velocity of all of the in-gap states, new states are extracted from the bulk and appended to the edge band to enable it to cross the bandgap. The way that this is achieved is closely related to the system's dimensionality: the degrees of freedom required to support a 1D wideband slow mode are extracted from the 2D reservoir of bulk states.

From the perspective of experimental implementation, the edge termination must be designed using the degrees of freedom that can be tuned in the underlying photonic topological insulator. This will require further studies aimed at adapting our proposal to specific photonic systems. In photonic crystals that break time-reversal symmetry through magneto-optics \cite{Haldane2008,wang2009observation}, one could, for instance, engineer an edge termination by varying the applied magnetic field in the vicinity of the edge or by further patterning the positions and radii of the edge sites. In proposals that utilize temporally modulated coupled resonators \cite{Fang2012,Minkov2016} or driven optomechanical cavities \cite{Fang2017}, both the coupling amplitudes and the pattern of relative modulation phases could be structured near the edge. Our work motivates further studies of how these degrees of freedom can be engineered along the edges of a system to control the properties of topological edge states and, in particular, to generate robust slow light.

\begin{acknowledgments}
M.C.R. acknowledges the National Science Foundation under award number ECCS-1509546, the Charles E. Kaufman Foundation, a supporting organization of the Pittsburgh Foundation, and the Office of Naval Research under YIP program, grant number N00014-18-1-2595.
\end{acknowledgments}
\bibliography{slowlightbib}

\onecolumngrid
\newpage
\begin{center}
\large
\textbf{Supplemental material: Broadband topological slow light through higher momentum-space winding}
\end{center}

\makeatletter
\renewcommand{\thefigure}{S\@arabic\c@figure}
\setcounter{figure}{0}

\twocolumngrid
\textit{Optimized edge terminations}.\textemdash Ideally, a slow-light mode should have a constant group velocity over its bandwidth so that different frequencies travel with the same speed and the structure of an input pulse is preserved through the course of propagation. For the edge termination introduced in the text, the resulting edge mode deviates from ideal linear dispersion and, in certain cases, even becomes non-monotonic. Here we describe a simple gradient descent optimization that enables us to significantly improve the dispersion of the edge mode.
 
We use the edge termination discussed in the main text as an initial seed for the optimization and seek to tune the couplings in a way that produces a more linear dispersion. In performing the optimization, we want to ensure that the Hamiltonian continues to respect translation symmetry parallel to the edge (i.e., for translation along $\mathbf{R}_1$). A simple way to achieve this is to consider the strip-geometry Bloch Hamiltonian $H(k_x)$ and define a set of optimization variables that rescale its non-zero entries. 

We take $H(k_x)$ to be the Bloch Hamiltonian defined for the standard zig-zag edge of the Haldane model, with no additional pattern applied to the edge termination. We then introduce a real symmetric matrix $\lambda_{ij}$ that rescales the entries of $H(k_x)$:
\begin{equation}\label{eqn_rescale}
\tilde{H}_{ij}(k_x) = \lambda_{ij}H_{ij}(k_x).
\end{equation}
The requirement that $\lambda_{ij}$ is symmetric is imposed to guarantee that the rescaled Hamiltonian is Hermitian. The resulting Bloch Hamiltonian $\tilde{H}(k_x)$ corresponds to a lattice model with rescaled first-neighbor and second-neighbor couplings and which respects translation symmetry along $\mathbf{R}_1$. We note that by rescaling the Bloch Hamiltonian as in Equation \ref{eqn_rescale}, we have, as a by-product, imposed a constraint in which distinct position-space couplings that enter into the Bloch Hamiltonian through the same $ij$ entry get rescaled by the same value. This corresponds to imposing a reflection symmetry about the $y$-axis on the rescaling factors. This constraint can be removed at the expense of expanding the set of optimization variables. However, the optimization results we obtain in the presence of this constraint are sufficient for our purposes.

We take the entries of $\lambda_{ij}$ to constitute our optimization variables. Since $\lambda_{ij}$ is a symmetric matrix, we assign only one optimization variable to both $\lambda_{ij}$ and $\lambda_{ji}$. Note also that if $H_{ij}(k_x)$ is identically zero, then multiplying by the corresponding entry $\lambda_{ij}$ has no effect. We, therefore, exclude such entries from our set of optimization variables. We define $v(k_x)=\frac{dE}{dk_x}$ as the group velocity of the bottom-localized edge mode of the rescaled Bloch Hamiltonian $\tilde{H}(k_x)$. In defining $v(k_x)$, we use an extended Brillouin zone scheme for $k_x$ so as to uniquely label the points of the edge mode when it winds many times around the Brillouin zone. We choose a reference group velocity, $v_0$, and perform an optimization to minimize the difference between $v(k_x)$ and $v_0$ at all $k_x$ for which the edge mode resides in an energy optimization window that we take to be equal to 70\% of the bulk bandgap (with the interval centered on $E=0$). We perform the optimization over this reduced window to allow for a region of variable group velocity in which the edge mode can transition to meet the bulk. We define $\mathcal{K}$ as the set of $k$-points, using the extended Brillouin zone scheme, for which the edge mode resides in this energy window (in the event that the mode exits and then reenters the window as $k_x$ is varied, $k$-points beyond the initial exit point are not included in $\mathcal{K}$). 

With the above definitions, we take our objective function, $S$, to be given by
\begin{equation}
S(\lambda_{ij}) =\frac{1}{2|\mathcal{K}|} \sum_{k_x \in \mathcal{K}} \left[\frac{v(k_x)-v_0}{v_0}\right]^2
\end{equation}
where $|\mathcal{K}|$ is the number of $k$-points contained in $\mathcal{K}$. The gradient of $S$ with respect to the $\lambda_{ij}$ can be computed using perturbation theory and is given by
\begin{equation}
\frac{\partial S}{\partial \lambda_{ij}} = \frac{\alpha_{ij}}{|\mathcal{K}|} \sum_{k_x \in \mathcal{K}} \text{Re}\Bigg[\frac{d\big ( H_{ij}(k_x)\mathcal{P}_{ji}(k_x) \big)}{dk_x}\Bigg]\Bigg[\frac{v(k_x)-v_0}{v_0^2}\Bigg]
\end{equation}
where $\mathcal{P}_{ji}(k_x) = \langle j|E(k_x)\rangle \langle E(k_x)|i\rangle$ is the projector onto the eigenstate $|E(k_x)\rangle$ of $\tilde{H}(k_x)$ (note the tilde) and we have introduced a combinatorial factor $\alpha_{ij}$ that accounts for the symmetry of $\lambda_{ij}$ and is defined to be 2 if ${i \ne j}$ and 1 if ${i = j}$.

Using the above gradient, we perform a numerical gradient descent to find a local optimum in the variables $\lambda_{ij}$. After each iteration of the algorithm, we update the set $\mathcal{K}$ since, during the course of optimization, states can enter and exit the energy window. We perform the optimization using a strip-geometry unit cell containing $200$ sites and using a discretization of the first Brillouin zone consisting of 40 $k$-points. We choose the initial values for the $\lambda_{ij}$ to produce the edge termination described in the main text (applied only to the bottom edge). We perform independent optimizations for different values of $n$ (see main text for the definition of $n$) defining the initial data.

\begin{figure}
\centering
\includegraphics[width=\columnwidth]{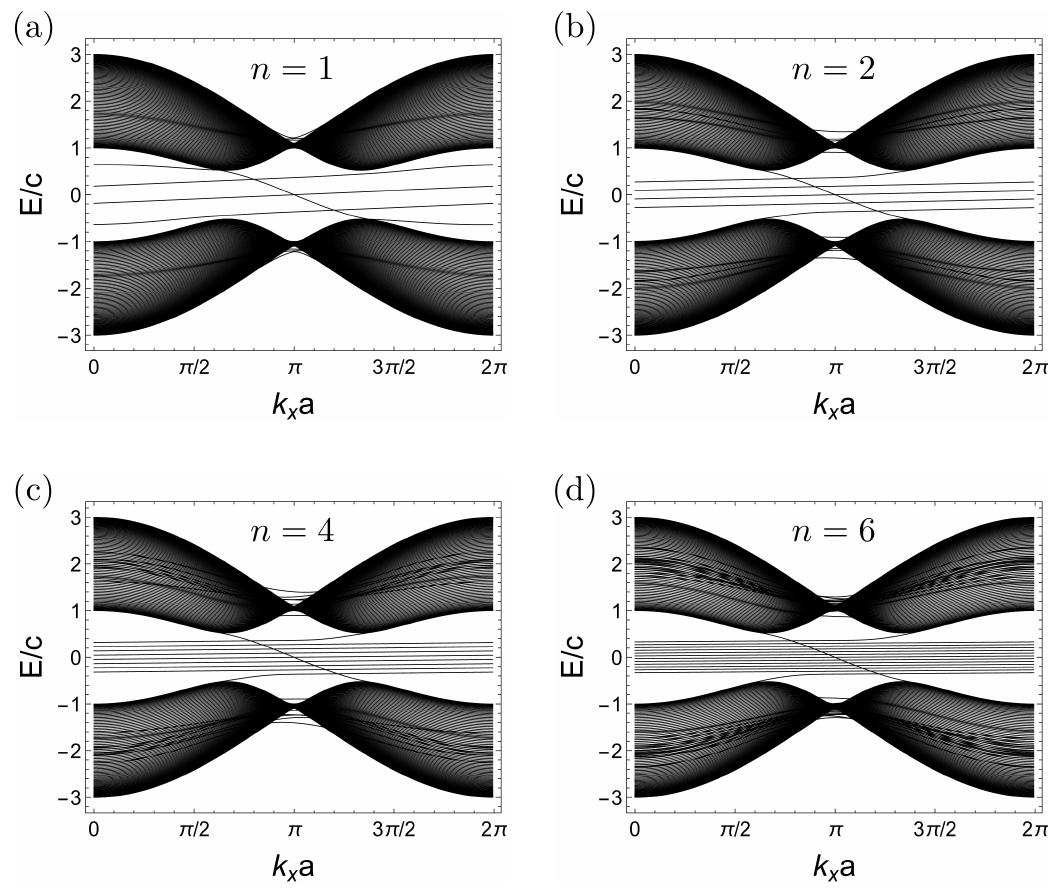}
\caption{\label{fig_optimized} Band structures for the optimized edge terminations. Panels (a)--(d) show the results obtained by using the $n=1,2,4,6$ edge terminations (described in the main text) as initial data for the optimization. For the optimized structures, deviations of the edge bands from ideal linear dispersion have been significantly reduced.}
\end{figure}

\begin{figure}[t]
\centering
\includegraphics[width=0.77\columnwidth]{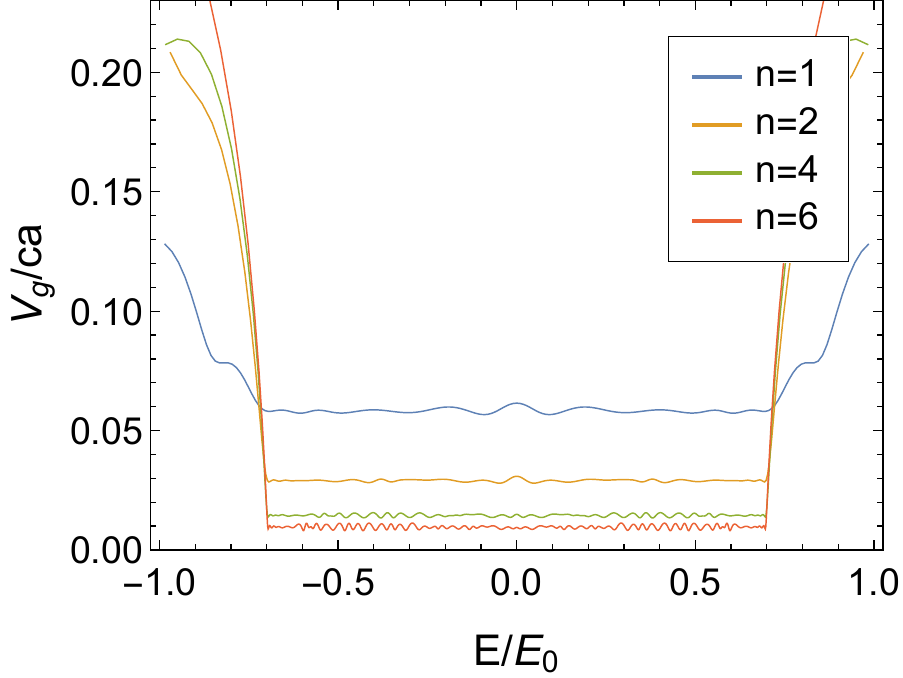}
\caption{\label{fig_groupvelocity}Group velocity for the bottom-localized optimized edge mode plotted as a function of energy. A separate curve is shown for each of the band structures of Fig. \ref{fig_optimized}. The energy, $E$, has been normalized to units of the bulk bandgap which spans $[-E_0,+E_0]$. Within the optimization energy window, $[-0.70E_0,+0.70E_0]$, the group velocity remains approximately constant, although small oscillations remain.}
\end{figure}

The band structures for the optimized terminations are shown in Fig. \ref{fig_optimized} for $n=1,2,4,6$ (the values used for the various plots of the main text). The resulting optimized edge bands are monotonically increasing and the deviations from ideal linear dispersion have been significantly reduced. While, as shown in Fig. \ref{fig_groupvelocity}, the group velocities of the optimized edge modes are approximately constant across the energy optimization window, there remain small oscillations not removed by the optimization. Further reduction of these oscillations may be achievable using a more complex optimization (e.g., a search for a global instead of a local optimum, or an optimization that tunes the phases of the couplings in addition to their magnitudes).

\textit{Time-domain simulations with disorder}.\textemdash Here we describe the details of the simulations presented in the main text in which we compare the behavior of slow light in the presence of disorder for topological and trivial systems. For both the topological and trivial cases, we consider a structure that begins with a fast-light region, transitions to a slow-light region, and ends with another fast-light region. For the topological system, the slow-light region is obtained using the optimized edge terminations described above and the fast-light region is obtained using an unmodified zig-zag edge. The use of the optimized structures is, in general, important (though not strictly necessary for the $n=1$ and $n=2$ cases at mid-gap) since it eliminates non-monotonicity of the edge dispersion that is present in some of the non-optimized structures and that introduces backward propagating states. The fact that the system is topological means that an optimization that eliminates all backward propagating states is, in fact, possible. In contrast, for a topologically trivial, time-reversal invariant system, backward propagating states will always be present. 

For the topologically trivial system, we use a 1D array with only nearest-neighbor coupling and vary the coupling strength to produce the fast-light and slow-light regions. In the trivial case, light incident on an abrupt interface between the fast-light and slow-light regions will be partially reflected. To avoid this, we introduce a linear taper spanning 20 lattice sites that adiabatically interpolates the couplings between the regions.

For the topological system, we excite the mid-gap topological edge state in the fast-light region and allow it to evolve through and exit the slow-light region. We take the initial pulse to have a narrow bandwidth, chosen to match the bandwidth of the pulse used in the 1D trivial array (a narrow bandwidth is necessary for the trivial system since its band structure narrows as the coupling constant is reduced). The slow-light region extends for a distance of $80a$, where $a$ is the lattice constant. We perform separate simulations using the $n=1$ and $n=2$ optimized edge terminations for the slow-light regions. Defining $v$ and $v'$ as the fast-light and slow-light group velocities, the $n=1$ and $n=2$ optimized edge terminations yield, respectively, $v/v' = 9$ and $v/v'= 18$.

For the trivial structure, we inject a rightward traveling pulse centered on $E=0$ that has a narrow pulse bandwidth contained within the bandwidth of the slow-light portion of the array. The length of the slow-light region is chosen to produce the same total pulse delay time as the topological case after accounting for the extra $20$-site adiabatic tapers at the fast-light and slow-light interfaces. We choose the coupling constants in the fast-light and slow-light regions so that the group velocities match those of the topological structure discussed above. To enable comparison of group velocities between the two structures, we assume the nearest-neighbor spacings for both the topological structure and the 1D array are equivalent.  As a result, matching the group velocities requires accounting for a factor of $\sqrt{3}$ difference in lattice constants (note also that distances in Fig. 4(a) and Fig. 4(b) of the main text are both normalized to the lattice constant $a$ of the topological structure in order to facilitate comparison). To match the fast-light group velocities, we take the nearest-neighbor coupling for the 1D array to be $0.48c$, where $c$ is the nearest-neighbor coupling of the bulk topological structure. Note that, in addition to matching the fast-light group velocities, this choice of couplings produces a bandwidth for the 1D array that approximately matches the maximum of the local (in $k_x$) bulk bandgap of the topological structure. For the slow-light region with $v/v'=9$, we take the coupling to be $0.053c$.  For the slow-light region with $v/v'=18$, we take the coupling to be $0.027c$.  

For both the topological and trivial structures, we add onsite disorder sampled uniformly from the interval $[-W,+W]$, with $W=0.05E_0$. Here $E_0$ is the scale of the bulk bandgap of the topological structure which spans $[-E_0,+E_0]$. The disorder is added to both the fast-light and slow-light regions. Note that for the topological case, the pulse has support on a 2D lattice (though it remains localized near the edge of the structure). To generate Fig. 4(a) of the main text, we have collapsed the 2D profile of the pulse over a strip-geometry unit cell (i.e., by summing the intensities of all the sites in the unit cell) in order to plot the pulse along its propagation direction. 

\end{document}